# The Labyrinth of Quantum Theory

## Tim Maudlin

Quantum mechanics predicts many surprising phenomena, including the two-slit interference of electrons. It has often been claimed that these phenomena cannot be understood in classical terms. But the meaning of "classical" is often not precisely specified. One might, for example, interpret it as "classical physics" or "classical logic" or "classical probability theory". Quantum mechanics also suffers from a conceptual difficulty known as the measurement problem. Early in his career, Hilary Putnam believed that modifications of classical logic could both solve the measurement problem and account for the two-slit phenomena. Over 40 years later he had abandoned quantum logic in favor of the investigation of various theories—using classical logic and probability theory—that can accomplish these tasks. The trajectory from Putnam's earlier views to his later views illustrates the difficulty trying to solve physical problems with alterations of logic or mathematics.

Introduction: The Siren Song of Quantum Theory

It has often been said that quantum mechanics is mysterious to us because it requires concepts that we cannot readily comprehend and displays behavior that cannot be understood in any classical way. And the term "classical" in this dictum has been taken in many senses. At the more trivial end of the spectrum, quantum theory is incompatible with classical Newtonian mechanics, which is not very surprising. But some claim that it is incompatible with a "classical" physical picture in a broader sense, for example demanding the amalgamation of the seemingly incompatible concepts of a particle and a wave. Classical probability theory has been fingered as the culprit, with quantum mechanics demanding negative probabilities. The "classical" idea that there must exist a complete description of the physical world has been questioned. And in the most extreme variation, "classical" logic has been called into question.

There has, from the beginning, been a penchant for *wanting* quantum theory to shock our whole conceptual system. Niels Bohr was fond of making gnostic pronouncements about the theory, and then touting their very incomprehensibility as an indicator of their profundity. He is quoted by his son as saying that there are "two sorts of truth: profound truths recognized by the fact that the opposite is also a profound truth, in contrast to trivialities where opposites are obviously absurd"[1]. The Theory of Relativity certainly required a substantial conceptual retooling, but it never provoked this sort of oracular commentary.

Are the phenomena associated with quantum theory intrinsically more difficult to account for than those associated with Relativity? It is said that the deepest mysteries of quantum theory are illustrated by the 2-slit interference of electrons. If an electron beam is sent through a single slit towards a fluorescent screen, there will be a somewhat fuzzy patch that lights up on the screen. Close that one and open a slit a bit to the right and the patch reappears on the right. But with both slits open alternating light and dark bands appear, the signature of the interference of waves. Turn the source down until only one spot at a time appears on the screen and the accumulated data still shows the interference bands. Individual electrons seem to somehow interfere with themselves.

These phenomena are referred to as particle/wave duality. Insofar as the interference bands emerge, the electron behaves like a wave. Insofar as individual local marks form on the screen, it behaves like a particle. What is it then? A wavicle? What does that mean?

Bohr's approach was to point to the experimental environment. In some circumstances the electron behaves like a particle, in others like a wave, and never like both. In itself it is both or neither. Classical concepts such as particle and wave, or position and momentum, only have content in particular experimental situations. Furthermore, those experimental situations may exclude each other, so at best only one of the two concepts can be deployed at a time. Which of the two is up to us. Thus the doctrine of complementarity was born. Our puzzlement about quantum theory arises from the habit of trying to use both concepts at the same time, as is done in classical physics.

This narrative, starting with experimental results and ending with conceptual revolution, appears to move smoothly from link to link. It certainly convinced several generations of physicists. So it can come as a shock to see how flimsy the whole construction is. Almost contemporaneously with the birth of quantum mechanics, Louis de Broglie suggested understanding the formalism in terms of particles guided by a "pilot wave", which was represented by the quantum wavefunction. De Broglie's theory was clear and comprehensible, and adequate to explain the phenomena. But it was roundly attacked and he abandoned it for many decades.

What was so objectionable about de Broglie's approach? Perhaps it was distasteful at the time due to its very comprehensibility. John Bell called such accounts of quantum phenomena "unromantic", as contrasted with the bracing radicalism of Bohr. He writes:

> The last unromantic picture I will present is the 'pilot wave' picture. It is due to deBroglie and Bohm. While the founding fathers agonized over the question
>
> 'particle' *or* 'wave'
>
> de Broglie in 1925 proposed the obvious answer
>
> 'particle' *and* 'wave'.
>
> Is it not clear from the smallness of the scintillation on the screen that we have to do with a particle? And is it not clear from the diffraction and interference patterns that the motion of the particle is directed by a wave?....This idea seems so natural and simple, to resolve the wave-particle dilemma in such a clear and ordinary way, that it is a great mystery to me that it was so generally ignored.[2]

What I hope this little historical prologue indicates is the seductiveness of the notion that quantum theory presents us with an even more radical break from "classical" patterns of thought and explanation than Relativity's break from the "classical" account of space and time. So seductive is the idea that rather plain and conceptually clear solutions to puzzles have been ignored while ever more outrageously baroque approaches are embraced. This pattern appeared—for a definite and delimited time—in Hillary Putnam's consideration of quantum theory. This is the story of that episode.

Beginnings: 1965

In 1965, Putnam published two papers touching on quantum theory: "Philosophy of physics"[3] and "A philosopher looks at quantum mechanics"[4]. Both of these pieces are notable for their clarity concerning the main problems confronting the theory. Putnam focuses almost entirely on what is called the measurement problem or the problem of Schrödinger's cat. The standard Copenhagen interpretation, associated with Bohr, is not a clear doctrine but it does have some constantly repeated themes. One is that there is no "anschaulich" or

"visualizable" picture of the microscopic realm that the theory addresses. "Classical" intuitions and expectations fail there. But in a curious twist, the macroscopic realm of everyday experience *must* be described in "classical" terms. Not via classical physics, of course: if classical physics could account for the behavior of macroscopic objects then it would be completely empirically adequate and there would be no call to adopt quantum theory. But the basic classical vocabulary of objects with both positions and velocities that move continuously through space must be used for describing both the configuration of experimental conditions and the outcomes of experiments. Those outcomes, in turn, must depend on the microscopic realm somehow. The main question was how to get a clear physical story about the relationship between these two domains.

John von Neumann had famously presented an axiomatized account of quantum theory, and in his account the interface between the microscopic realm of atoms and the macroscopic realm of everyday life took a particularly shocking form. Von Neumann postulated two completely different dynamical equations for the wavefunction that is used to represent the physical state of a quantum system. So long, von Neumann said, as the system was not *measured*, its evolution was the smooth, deterministic change given by Schrödinger's equation. But when the system in intervened upon by a *measurement*, it suddenly and unpredictably jumps into one of the eigenstates associated with the operator that represents the measured quantity. The chances for the different possible outcomes are given by Born's rule.

Here are von Neumann's own words:

We therefore have two fundamentally different types of interventions that can occur in a system **S** or in an ensemble [**S**$_1$,...., **S**$_N$]. First, the arbitrary changes by measurements which are given by the formula (**1**.) [here von Neumann writes the transition from a single initial state to a statistical mixture of final eigenstates, each with a probability]....Second, the automatic changes that occur with the passage of time. These are given by the formula (**2**.) [Here von Neumann gives the Schrödinger evolution].[5]

For all of its seeming mathematical rigor, von Neumann's scheme is, as Bell would say, unprofessionally vague. Surely a measurement is just some sort of physical interaction between systems just as otherwise occurs with the passage of time. After all, what *physical* characteristic could distinguish measurements from other sorts of interactions, a characteristic so precise that its presence or absence makes the difference between smooth deterministic evolution and stochastic jumpy "collapses". This, in a nutshell is the measurement problem.

The Putnam of 1965 was rightfully scandalized by the state of the foundations of quantum theory. He finds the Copenhagen approach unsatisfactory as it seems to imply that particles only have positions and momenta when they are measured and not in between measurements.[6] Furthermore, and rather ironically, this interpretation is forced to say that the physical procedures we call "making a measurement" are really no such thing: they do not *reveal* some pre-existing property of the target system but rather serve to *create* a new property. Because of these strange features, Putnam discusses so-called "hidden variables" theories in which there is more to the physical state of a system than is reflected in the wavefunction. The de Broglie/Bohm 'pilot wave' theory was even then the most famous example, and Putnam discusses it in detail. He says that he doesn't like it because it is mathematically contrived, and also raises a problem that, oddly, requires applying classical physics to a quantum-mechanical situation.

---

5 John von Neumann, *Mathematical Foundations of Quantum Mechanics* (Princeton: Princeton University Press, 1955), 351.
6 Putnam, *Mathematics, Matter and Method*, 141.

Putnam also discusses Eugene Wigner's attempt to tie collapse of the wavefunction to human consciousness. [7] He finds all such approaches that take the collapse to be a physical event—rather than just an updating of subjective probabilities on receipt of new information—to be unacceptable, concluding that

> *no* satisfactory interpretation of quantum mechanics exists today. The questions posed by the confrontation between the Copenhagen interpretation and the hidden variable theorists go to the very foundations of microphysics, but the answers given by the hidden variable theorists and the Copenhagenists are alike unsatisfactory.[8]

What is notable about these two examinations of quantum theory from 1965 is that the question of logic is never raised. The next time Putnam weighs in, the situation is quite different.

The Quantum Logic of David Finklestein

---

[7] Putnam, *Mathematics, Matter and Method*, 147.
[8] Putnam, *Mathematics, Matter and Method*, 157.

In 1968 Putnam returns to quantum theory with "The logic of quantum mechanics".[9] As the title indicates, questioning—or at least discussing—logic itself is now on the main menu. Our quick review of the 1965 works have set the table: in those, Putnam rejects real physical collapses of the wavefunction, "hidden" variables, and the Copenhagen interpretation because none of them could deliver what he regarded as a plausible account of measurement as a purely physical interaction between physical systems. The question that we must keep foremost in our minds is how the introduction of a new logic, or even more radically the replacement of classical logic with a new logic, could even in principle help to resolve this problem.

The main conceptual issue here concerns the question of what *empirical* import the logical connectives in a language can have. Let's take & as a clinical example. In propositional logic, & is a sentential connective: two well-formed formulas connected by & constitute another well-formed formula. The "meaning" of the connective can be given by a truth table or by a set of valid inferences. If every well-formed formula is either true or false, then the truth table is simple to specify: A & B is true if both A and B are true and is false otherwise. In terms of inferences, we give &-Introduction and &-Elimination rules. &-Introduction says that from A and B as premises one can derive A & B as a conclusion. &-Elimination says that from A & B one can derive A and one can derive B. It is easy to see that the validity (truth-preservingness) of &-Introduction requires that the truth of both A and

<hr>

[9] Reprinted as Chapter 10 of *Mathematics, Matter and Method*, 174-197.

B be sufficient for the truth of A & B, and &-Elimination requires that it be necessary.

Now if one were to reject bivalence, and allow well-formed formulas to be other than True or False, then the semantics of & would have to be revisited. If this new scheme retains the old inference rules, then again the truth of both A and B would still have to be necessary and sufficient for the truth of A & B. But if the truth values expand beyond "true" and "false" then the conjunctions that include the new values have to be addressed.

But what possible *empirical* consequences could such a change in logic have? How could anything like this solve the *physical* issues that must be confronted when trying to make clear sense of quantum theory? That is the main question we must bear in mind in the sequel.

The Putnam of 1965 thought something had to be done, but did not know what. By 1968, the idea that logic had to be amended came forward, but not because Putnam's own investigations had led him in that direction. Rather, as he makes clear, he was brought to the approach by David Finkelstein. Finkelstein, in turn, was working in a tradition that goes back to von Neumann and Garrett Birkhoff.[10] The mathematical core of this approach adverts to the structure of subspaces of Hilbert space.

Hilbert space is the space of all quantum-mechanical wavefunctions. It is an infinite dimensional space, essentially just because wavefunctions are *functions*, and a function is specified at an

---

infinitude of points. The other piece of mathematical machinery we need are *Hermitian operators*, also commonly called *observables*. These operators, as their name suggests, operate on wavefunctions: hand an operator a wavefunction and it hands another back to you. In special cases if you hand the operator a wavefunction it hands back the very same wavefunction multiplied by a number. Such a wavefunction is called an *eigenfunction* of the operator and the number is called its *eigenvalue*.

At this point, a key physical assumption is made. The "observable" operators are associated with *physical magnitudes* and the condition for a system to have a definite physical magnitude is that its wavefunction be an eigenfunction of the associated operator. The value of the physical magnitude is then its eigenvalue. Any wavfunction that is *not* an eigenfunction of the operator *fails to have any value for the magnitude at all*. Thus, a wavefunction that is not an eigenfunction of the position operator represents a system that fails to have any definite position, and a waverfunction that is not an eigenfunction of the momentum operator represents a system that fails to have any definite momentum. And since no position eigenfunction is a momentum eigenfunction, mathematics alone guarantees that no system can simultaneously have a definite position and a definite momentum. Indeed, the closer a system comes to having a definite momentum, the further it is from having a definite position and vice versa. This is an instance of the *uncertainty relation* or *complementarity.*

The final piece of the technical apparatus arises from the observation that the eigenfunctions of an observable that have the same eigenvalue form a *subspace* of the Hilbert space, called its eigenspace. So the condition for a system to have a particular value of an observable is that its wavefunction lies in that eigenspace. Therefore every proposition of the form "System S has value V for observable O" can be associated with a particular subspace of Hilbert space. The truth condition for the proposition is just the system wavefunction lying in the subspace. And the falsity condition is being perpendicular to the subspace. If S's wavefunction is neither in nor perpendicular to the eigenspace of the value V then the proposition is neither true nor false, and a measurement of the observable might or might not yield the value V.

Let's pause for a moment to think about this last case. There is some observable O and the wavefunction for the system S does not lie in any eigenspace of O, so the quantum predictive apparatus makes only a probabilistic prediction for what would happen were you to measure O. What options are there for assigning a truth value to the claim that S has a particular value of the quantity O?

One thing one might say is that if, in fact, S has the value V then, by the definition of what a "measurement" is, a measurement would have to return the outcome "V". And since there is no value that the measurement is certain to return if it should happen to be done, there is no value that S has.

But this chain of reasoning has several flaws. One is that what we usually call "measurements" just don't have the feature mentioned. If I put a thermometer into a liquid to "measure" its temperature, the reading is typically not the temperature that the liquid had before I

"measured" it. If the thermometer was colder than the liquid then the equilibrium temperature of the liquid-thermometer system at the end is lower than the original temperature of the liquid, and if the thermometer was hotter then it is higher. Maybe one considers the discrepancy unimportant, or maybe it is important enough to try to compensate for. But to do that, one has to understand in complete physical detail how the "measurement device" physically interacts with the "system" to yield and outcome. And that, recall, just *is* the measurement problem! So on this approach one can't really know how to assign truth values to propositions without solving the measurement problem. To appeal to a new logic to solve the problem would be a viciously circular procedure.

The second problem is that the fact that quantum mechanics only generates probabilistic predictions about the outcomes of measurements does not, by itself, imply anything about the systems themselves. It might be that the dynamics of the system is fundamentally indeterministic, and the system is neither disposed to yield a particular value nor not to yield it. Or it could be that the probabilities, and consequent statistical predictions, only reflect an inadequacy in the quantum formalism, and that the wavefunction used to make the predictions is, in Einstein's terminology, incomplete. That is, the wavefunction fails to specify all of the physical features of a system, and the statistical predictions arise because different systems assigned the same wavefunction are physically different. Among these differences could be the value of the observable quantity O. In that case, the fact that quantum mechanics makes no definite prediction for the outcome of an O measurement would not imply that S fails to have any value for O. To understand whether it does or does not again requires solving the measurement problem.

So there are two essential questions that the standard presentation of quantum mechanics fails to answer: 1) Is the wavefunction a complete physical description of a system or not? And 2) How does the pre-interaction-with-a-"measuring"-device state of a system relate to its post-interaction state? What does the outcome of the measurement indicate about the system before it was measured? Without answers to these questions—physical answers—the truth value of many claims cannot even be addressed. The quantum logic of Finkenstein purported to address them.

The key to Finkelstein's approach lies in the observation that there are three purely mathematical operations on subspaces of a space that bear some formal resemblance to the operation of the classical truth-functional operators. Given two subspaces of a vector space one can form their *intersection*, which is just all the elements of the space that lie in both of the subspaces. The intersection is itself a subspace. And given two subspaces one can form their *span*, which is just the set of all vectors that can be formed by adding vectors from the two subspaces. The span, too, is a subspace of the vector space. Finally, given a subspace of a vector space one can form its *orthocomplement*, which is just all of the vectors orthogonal to the subspace. It is yet another subspace.

So if we have associated every subspace of a vector space with a proposition, the intersection and span operations are operations on a pair of propositions that yield a proposition, and the orthocomplementation operation is an operation on a single proposition that yields a proposition. How are these various propositions related to each other in terms of their truth values?

Clearly, if proposition A is true because the state of S lies in the A-subspace, and proposition B is true because the state of S lies in the B-subspace, then the proposition associated with the intersection of the two subspaces, call it "A⊓B", is true because the state of S lies in the intersection of the subspaces. (This notation is not standard, but it will help keep things straight.) And if either A or B is anything other than true, then A⊓B cannot be true, because if a state does not lie in a space it cannot lie in any subspace of that space. So ⊓ functions just like the classical & with respect to true and false sentences.

Let's indicate the proposition associated with the orthocomplement of the A-subspace "⊥A". Obviously, if A is true then ⊥A is not true, and if ⊥A is true then A is not true. Again, if the only truth values we had were true and false, ⊥ would operate just like the classical negation. But here the divergence from classical negation is more severe. Since there are vectors that lie neither in the A-subspace nor in the ⊥A-subspace (assuming neither is the whole space), there are propositions such that neither it nor its orthocomplement are true. So ⊥ does not work like classical ∼ in the sense of mapping anything other than a true proposition to a true proposition.

Finally, what about the span? We will indicate the proposition associated with the span of the A-subspace and the B-subspace as "A⊔B". One the one hand, if the state of a system lies in the A-subspace or in the B-subspace then it lies in the span of the two. So if A is true or B is true then A⊔B is true. But there are many vectors that lie in the span of the A-subspace and the B-subspace but don't lie in in either. If the state of a system is one of these vectors, then A⊔B is true even though neither A nor B is true. So although ⊓ functions just like &, ⊔ is quite unlike the classical ∨.

Interestingly, despite these differences, many tautologies of classical logic remain tautologies (i.e. propositions true for every state) when the corresponding operators are substituted for each other. Suppose, for example, we just *define* the falsehood of A in the new system as the truth of ⊥A. Then there are some propositions in the new language that are neither true nor false: we need a third truth value. And, mixing the different sets of operators, for any such proposition A, A ∨ ⊥A is not true. But still, A ⊔ ⊥A is true for all A since A ⊔ ⊥A is associated with the entire space.

Not all of the inferences that are valid in classical logic remain valid when the new operators are substituted. For example, in classical logic A & (B ∨ C) implies (A & B) ∨ (A & C) and vice versa. But in the new system A ⊓ (B ⊔ C) can be true while (A ⊓ B) ⊔ (A ⊓ C) is false. The analog of the distributive law fails. (To see an example, let C be ⊥B, and let the system state lie in the A-subspace, but neither in the B-subspace nor the ⊥B-subspace. Then A ⊓ (B ⊔ ⊥B) is true, but (A ⊓ B) ⊔ (A ⊓ ⊥B) is false.)

The structure of subspaces of Hilbert space under the operations of intersection, span, and orthocomplementation in this way becomes isomorphic to the structure of the lattice of associated propositions under the operations of meet, join, and orthocomplement. That, in technical terms, is what Finkelstein's "quantum logic" is.

It is here that we reach a critical point in the discussion. There are two crucial questions. First, are there any grounds for thinking of the operators ⊓ and ⊔ and ⊥ on these propositions as *alternatives* rather than *supplements* to the classical truth-functional operators &, ∨, and ∼? And second, how could the induction of these new operators, whether as alternatives or as supplements, help us solve any interpretational problem of quantum theory?

Finkelstein's Logic and Understanding Quantum Theory

The most immediate question to ask when presented with something called "quantum logic" is whether it is being offered as a *supplement to* or a *replacement for* classical logic. If we don't know this, then we have no idea how to proceed.

If the new logic merely supplements the old, then adding the new logical connectives creates a more effective and extensive logical language, but does not revise any results of the old logic. All the old tautologies are still tautological, all the old valid inferences still valid. In this spirit, we could add ⊓ and ⊔ and ⊥ to a formal language alongside & and ∨ and ∼. If the vector representing a system falls in the span of the subspaces A and B, but not in the A-subspace and not in the B subspace, then we could truly write: (A ⊔ B) & ∼(A∨B). In such a way, we could convey more using the new language than we could have in the old.

Such a change might be a *useful innovation* in logic, but certainly not a *conceptual revolution*. There would have been nothing strictly *wrong* in the old language, which remains as a fragment nestled inside the new. At worst, we might have been tempted to incorrectly use ∨ instead of the more appropriate ⊔, or ∼ rather than ⊥.

Putnam's thought about quantum logic, though, is much more radical. The radical idea is that the new logic is to *replace* the old one rather than supplement it, and the formal similarity between ⊔ and ∨, ⊓ and &, and ⊥ and & makes clear how the substitution goes. In some sense disjunction *really is* ⊔ rather than ∨, and so on.

The key analogy here is supposed to be that of Euclidean and non-Euclidean geometry. For millennia, humans believed that they lived in a three-dimensional Euclidean space. Furthermore, they believed that they had some sort of *a priori* access to the structure of that space, codified in the theorems of Euclidean geometry. But eventually, empirical and explanatory pressures made it rational to supplant the Euclidean hypothesis with a non-Euclidean one. This was obviously a replacement rather than an enhancement: the old system is abandoned in favor of the new. One understands why the old worked as well as it did, but in the end it simply is regarded as a false approximation to the truth.

The analogy between geometry and logic is made explicitly in "How to think quantum-logically" (1974),[11] and is very telling. In the case of geometry, we definitely have a replacement rather than an expansion. The Euclidean geometry must die so that the non-Euclidean can live. And in the process, some empirical phenomena—such as the anomalous advance of the perihelion of Mercury—get explained.[12] Our main questions, then, are what could it even mean to *replace* classical logic with some other logic? And if we can get clear on that, how could any such replacement help solve any empirical problems?

---

[11] Hilary Putnam, "How to Think Quantum-Logically", *Synthese* **29** (1974), 55-61.

It is useful to have an example here. What empirical problems did the switch from Special Relativity (flat space-time geometry) to General Relativity (curved space-time geometry) help account for? The anomalous advance of the perihelion of Mercury, the bending of light passing near the Sun, the gravitational Red shift...these are the standard answers. None of these phenomena are logically incompatible with Special Relativity. Adjustments to overall physics elsewhere, that kept the flat space-time, are possible. But General Relativity accounts for them in a simple and satisfying way.

So what is the parallel example of a switch to quantum logic making the explanation of a physical phenomenon simpler and more elegant?

Interestingly, there isn't one. Instead, Putnam tries to argue that *classical* logic yields bad predictions, but not at all that *quantum* logic corrects these predictions. And the example he discusses is the famous two-slit interference effect. Let's quickly recap that.

If we fire elections or photons at a barrier with a thin slit, an image appears behind the barrier that is roughly the shape of the slit, but spread out in the direction that the slit is thin. It is brightest in the middle and shades off to either side. Let's call this the image formed when only Slit A is open. If we close Slit A and make a nearby parallel Slit B, the image just moves to behind Slit B. No surprise.

---

[12] Properly speaking, General Relativity does not so much demand replacing Euclidean with non-Euclidian geometry, as replacing flat Minkowski space-time with a variably curved space-time. But the analogy is very tight.

But what if both slits are opened? Then instead of just getting the sum of the two images, one gets a highly striped set of interference bands, rapidly alternating between bright and dark. The pattern is reminiscent of the interference patterns in water waves in similar circumstances. But curiously, the pattern persists even when the electrons are sent one at a time, and the images form one discrete spot at a time.

Now this behavior may be surprising, but Putnam makes a much, much stronger claim: it is not the behavior that *classical logic* leads us to expect! How does that argument go?

Putnam makes the following "classical" argument. If you want to know how likely it is for an electron to land in a particular region R on the screen (call the proposition that it lands there 'R'), first close slit B and measure the number with only A open. Call the proposition that the particle passes through slit A 'A'. So we get a number: prob(R/A). Do the same by closing slit A and opening B to record prob(R/B), the chance of hitting R is the particle goes through B. Now what about with *both* slits open? In that case, says Putnam, the particle must either go through Slit A or through Slit B, so for every electron that makes it to the screen we have the truth of A ∨ B. What we want is Prob(R/A ∨ B). Now by some fancy manipulations, Putnam "derives" that classical logic requires that Prob(R/A ∨ B) = ½ Prob(R/A) + ½ Prob (R/B). The appearance of the ½ comes as a bit of a shock: offhand, one would suppose that the chance of getting to R by going either through A or B is just the *sum* of the chance going through A with the chance going through B. But let's leave that curiosity aside. What a moment's thought reveals is that, whatever Putnam has calculated, it can't *possibly* be a consequence of "classical logic" or any other kind of logic!

Consider the situation. We do one experiment with one slit open. We record a number. We do another experiment with the other slit open. We record another number. Now we wonder what will happen with *both* slits open. Well, as far as logic of any kind goes, *anything* could happen! The apparatus could blow up. All of the electrons could be reflected back to the source. The electrons could all turn into rabbits. *Logic* won't prevent it!

The fallacy, once you point it out, is glaringly obvious. The first experiment shows how many electrons get through slit A and land in R *when only slit A is open*. And the second how many get through slit B and land in R *when only slit B is open*. From the point of view of logic, this tells us exactly nothing about what will happen *when both slits are open*. Formally, let's let O(A) stand for only A is open, O(B) stand for only B is open, and O(A,B) stand for both A and B, and only A and B are open. Then the first experiment measures prob(R/A & O(A)), and the second measures prob(R/B & O(B)). And what the two-slit experiment measures, assuming the electron goes through one slit or the other, is prob(R/(A ∨ B) & O(A,B)). But since the first experiments carry no information at all about the condition O(A,B), no logic in the world, of any kind, will allow you to derive the third probability from the first two.

Further, all Putnam attempts to show is that classical logic yields a bad prediction, not that quantum logic yields a good one. When we replace the space-time of Special Relativity with that of General Relativity, we not only abandon the Special Relativistic predictions, we acquire the new General Relativistic ones. If Putnam had tried to make a parallel argument, it would have become obvious that the first, classical, argument was invalid.

How could Putnam have been led so far astray? In "The Logic of Quantum Mechanics" he repeatedly emphasizes that the account of quantum logic is not his own: it is Finklestein's. His account lies somewhere between reporting and endorsing. He is perhaps trying to set the large-scale stage for the very idea of logic being empirical rather than pushing a particular account of the right logic. Of course, since the example fails so spectacularly, we are left with no clue about how logic really could be empirical. But Putnam was not done with the topic yet.

Gestures Towards Yet Another Quantum Logic

In 1974 Putnam returns to the question of quantum logic with the very short paper entitled "How to Think Quantum-Logically". As already mentioned, this paper starts with the proportionality [Euclidean] Geometry: General Relativity:: [Classical] Logic: Quantum Mechanics. I have supplied the bracketed clarifications, which are clearly intended. Once again the overarching theme is that just as General Relativity demands the rejection of Euclidean geometry as the price for a tremendous simplification of physics, so Quantum Mechanics offers us an elegant account of physical phenomena at the price of replacing classical logic. The paper is written in a slightly coy way. Putnam begins by expositing what "the advocate of the quantum,-logical interpretation of quantum mechanics claims", [13] as if this were a dispassionate presentation of one approach to understanding quantum theory. But he soon enough slips into the revelatory locution "we advocates of this interpretation"[14] indicating that a defense of quantum logic rather than merely an exposition is in store.

We have already seen that Putnam's attempt to tie the two-slit experiment to some logical principle failed. In this paper, no such empirical phenomenon as the two-slit experiment is even mentioned. And in a further surprise, there is no trace of Finklestein's quantum logic. Putnam is now setting out to create his own new logic, based on new arguments.

Fineklestein's logic, founded on the structure of subspaces of Hilbert space under the operations of intersection and span, made no change to the conjunction-like operation. The classical conjunction of two propositions is true iff the system's state lies in the intersection of the two propositions' subspaces. The novelty in Finkelstein's logic was all in the analogs of disjunction and negation. But Putnam's new logic is just the opposite. The novelty lies in the analog of conjunction! How is that supposed to work?

There are really two tricks that Putnam pulls here, one with disjunction and the other with conjunction. These tricks are both in

---

service of one main contention: that the interactions called "measurements" really deserve the name. That is, a "measurement" merely reveals a pre-existing fact about a system, rather than playing a role in *creating* some new fact. The very first characteristic of what Putnam calls the "quantum logical view of the world" is:

> (1) *Measurement only determines what is already the case: it does not bring into existence the observable measured, or cause it to 'take on a sharp value' that it did not already possess.*[15]

There are two ways to understand this fundamental principle. One is merely as a stipulation, as a *definition* of what it takes to count as a "measurement" in the sense Putnam wishes to use that term. As such, it would have nothing to do with logic at all, because "measurement" is not a logical particle.  And if this is just a stipulation, it is an open question whether, according to some theory, there are any "measurements". Since typical measurements are interactions, it is a question of theoretical analysis whether the "outcome" of the supposed "measurement" accurately indicates some pre-existing state of affairs.

Putnam evidently wants to take the term "measurement" a different way. He wants to regard certain experimental conditions as constituting paradigm "measurements", especially "measurements" of position and momentum. Having fixed that, principle (1) becomes a *constraint on the physical analysis of these paradigmatic "measurements"*.

---

[15] Putnam, "How to Think", 57.

The Putnam-style quantum logician is simply not open to a physics that analyzes these paradigm situations so that the outcome of the physical interaction between the system and the apparatus does not reveal a pre-existing physical magnitude. It is an odd thing to adopt such an *a priori* demand on these particular experimental arrangements. Surely *logic* cannot demand it! And even more than that, proofs due to Kochen and Specker,[16] Bell[17], and Greenberger, Horne and Zeilinger[18] show that *no* physics can possibly meet the requirements of principle (1). The principle defies not classical logic but mathematics itself if we want to recover the quantum-mechanical predictions. It is simply off the table.

If one demands that an acceptable physics meet a condition that is mathematically impossible to meet there is bound to be trouble. And basically Putnam's new logic tries to defuse that trouble by abandoning classical logic. Let's trace the steps.

First, Putnam takes a page out of Finkelstein. There are eigenstates of the position operator, and according to the interpretation Putnam is using the only situation in which a particle has a particular position is when it is in a eigenstate with that eigenvalue. And there are many, many states that are not eigenstates of the position operator.

---

[16] S. Kochen and E. Specker, "The Problem of Hidden Variables in Quantum Mechanics", *Journal of Mathematics and Mechanics*, **17** (1967), 59–87.

[17] J. S. Bell, "On the Einstein-Podolsky-Rosen paradox", *Physics* **1** (1964), 195-200.

[18] D. M. Greenberger, M. A. Horne and A. Zeilinger, "Going beyond Bell's Theorem," in *Bell's Theorem, Quantum Theory and Conceptions of the Universe*, M. Kafatos (ed.), Dordrecht-Boston-London: Kluwer (1989), 69–72.

Indeed, all but a set of measure zero are not position eigenstates but rather superpositions of position eigenstates. So if we are using classical logic it is correct to say than in most states the particle has no particular position.

But the *span* of all the position eigenstates is the whole Hilbert space, which includes every possible wavefunction. Similarly, the set of momentum eigenstates is a set of measure zero in the Hilbert space, but the span of that set is the whole Hilbert space. Putnam now considers a single particle, which, he says, has a position and has a momentum:

> In symbols, taking 'Oscar' to be the name of one of these particles:
>
> > (*Er*) the position of Oscar is *r*. (*Er'*) the momentum of Oscar is *r'*.
>
> But it must not be concluded that each of these particles has a position *and* a momentum! For the following is *rejected*:
>
> > (*Er*)(*Er'*)(the position of Oscar is *r*. the momentum of Oscar is *r'*).[19]

Since these two sentences are logically equivalent in classical logic, quantum logic must depart from classical logic. But how?

Finklestein's logic has the logical inequivalence of the two sentences because a "disjunction" (i.e. span) of two propositions can be true without any of the "disjuncts" being true. The existential claim is just an infinitary disjunction, so for Finklestein it can be true without any instance being true. So each conjunct in the first conjunction is true—even necessarily true—making the conjunction necessarily true as well. But the second sentence, instead of being a conjunction of (infinite) disjunctions, is an (infinite) disjunction of conjunctions. In Finklestein's logic, each of the conjunctions is false, since the intersection of a position eigenstate and a momentum eigenstate is zero. Since every conjunction is false, so is their infinite disjunction. It is true on Finklestein's logic that every particle has a position and every particle has a momentum, but no particle has both a position and a momentum. At least if we translate the span into English as if it were a disjunction.

---

19 Putnam, "How to Think", 57.

Putnam's quantum logic cannot achieve the inequivalence of the two sentences in the same way that Finklestein's does. Putnam's requirement that "measurements" merely reveal pre-existing physical facts demands that such physical facts *always* exist. Individual particles do not merely  have position-in-general without any specific position, as Finklestein would have it, but they always have some particular position just as the classical disjunction would demand. And similarly for Putnam's particle's momentum: unlike Finklestein's, it has not merely momentum-in-general but always a particular (albeit unknown) momentum. For Putnam, unlike for Finklestein, the existential sentences are true because some specific instance is true. The semantics of disjunction is just classical semantics. And the conjunction of the existentials is true because each conjunct is true.

The pressing question, then, is how Putnam intends to achieve the semantic difference between the conjunction of the existentials and the double existential over the conjunction. If Oscar really has some specific position and some specific momentum, how can Oscar fail to have both a specific position and momentum?

Putnam bites the only bullet he has left: the problem with classical logic is not disjunction, but conjunction! In classical logic, the truth of the conjuncts separately implies the truth of the conjunction. But for Putnam, each of the conjuncts can be true without the conjunction being true. This is not a possibility that Finklestein ever suggested.

But how can that be? How can A be true and B be true but A & B fail to be true? What does that even mean? Are there any vaguely analogous examples of the failure of conjunction in an everyday context?

Unfortunately, Putnam never directly confronts this issue. He simply says that the conjunction of a particular position and a particular momentum cannot be true, since otherwise we would not have the principle of complementarity! Here is the relevant passage:

> Complementarity is fully retained. For any particular $r$ and any particular $r'$ the statement 'Oscar has position $r$ and Oscar has momentum $r'$ ' is a logical contradiction. It is, of course, just the sacrifice of the distributive law that we mentioned a few moments ago that enables us to simultaneously retain the objective conception of measurement as finding out something which exists independently and the objective conception of

> complementarity as a prohibition on the simultaneous existence of certain states of affairs, not just as a prohibition on simultaneous knowledge.[20]

But the distributive law has really nothing to do with the case. If position measurements always reveal pre-existent positions and momentum measurements always reveal pre-existent momenta, then the particles always have pre-existent positions and pre-existent momenta. One need not appeal to any distributive law to conclude that Oscar always has both.

What Putnam has in mind is only revealed on the last page of the article. There he asserts—contrary to both Finklestein and to quantum theory—that a system does not have one wavefunction ("state vector") but *many*. Since no single wavefunction can ascribe both a position and a momentum to Oscar, and Putnam want to maintain that Oscar always has both a position and a momentum, he is forced into this position. But somehow, although Oscar is described by both of these state vectors simultaneously, one is not allowed to *say* so! For that, according to Putnam, would be a quantum-logical contradiction.

If this situation sounds incoherent, that's because it is. Here is part of the next to last paragraph of the paper:

> Thus we get: a system *has* more than one state vector, on the quantum-logic interpretation, but one can never *assign* more than one state vector! Or, to drop talk of 'state vectors' altogether…we may say: A system has a position *and* it has a momentum. But if you *know* the position (say *r*),

---

[20] Putnam, "How to Think", 58.

you cannot *know* the momentum. For if you did, say, know
that the momentum was *r'*, you would know 'Oscar has the
position *r*. Oscar has the momentum *r''*, which is a logical
contradiction.[21]

It is really impossible to make sense of this paragraph.

First, the mention of *knowledge* comes out of nowhere: up until
now, no epistemic issues have even been raised. But further, if all
propositions of the form 'Oscar has the position *r*. Oscar has the
momentum *r''*' are logical contradictions, then if I know Oscar has
position *r*, it is not just that I can't *know* Oscar's momentum, but Oscar
can't *have* a momentum. Whether I know it or not is neither here nor
there. So Putnam's desire for Oscar to have both a definite position and
a definite momentum collapses.

Clearly, Putnam's new original quantum logic was a work in
progress at this point. In his last paper on this topic, he tries to clarify
the situation.

Quantum Logic as the Logic of Prediction

Putnam's attack on the "logic" of conjunction finally gets a
coherent explication in the 1981 paper "Quantum Mechanics and the
Observer".[22] Once again, all of our foundational questions arise. Is
"quantum conjunction" supposed to *replace* classical conjunction or

---

[21] Putnam, "How to Think", 61.

[22] Hilary Putnam, "Quantum Mechanics and the Observer", *Erkenntnis* **16**, no. 2 (July 1981), 193-219.

*supplement* it? And what could possibly go *empirically* wrong with the classical truth-functional operator? If one just stipulates that A & B is true iff A is true and B is true, how could such a stipulation lead to any new empirical consequences at all? There certainly seems to be no bearing of conjunction on the two-slit experiment, where at least the issue of disjunction ("the particle goes through slit A or the particle goes through slit B") appears to be relevant to any proposed explanation. What issue is there with conjunction?

We last left Putnam in a bind. On the one hand, he wants the experiments we call "measurements" to just reveal some pre-existing physical property of a system rather than play a causal role in creating the property. And since position measurements and momentum measurements always have outcomes, it follows that particles always have both positions and momenta. But according to the Finklestein quantum logic, the conjunction of a proposition assigning a particle a position and a proposition assigning the particle a momentum is *logically false*. No state vector assigned to the system can make both true. What to do?

In "Quantum mechanics and the observer", Putnam doubles down on the assertion that particles always have both position and momentum. This is true even though position (at a time) and momentum (at that time) are quantum-mechanically incompatible observables. Even more than that, he asserts that sometimes we can *know* both of these two incompatible observables to arbitrary accuracy, so long as the knowledge is about some time in the past. Suppose, for example, there is a source with a shutter in it surrounded by a spherical screen. If the shutter is opened briefly at $t_0$, resulting in a spherical wavefunction propagating outwards, and a flash is observed in some small region R of the screen at $t_1$, then the experiment tells us both that a particle escaped at $t_0$ and that it was in R at $t_1$. But, as Putnam points out, the quantum mechanical states associated with these properties are incompatible: one cannot prepare any state that is certain to both pass through the open shutter and arrive the appropriate time later at R. Each of these properties has an associated quantum state, $|\psi_0\rangle$ and $|\psi_1\rangle$, but these states are incompatible. Their quantum-logical conjunction is necessarily false. Nonetheless, says Putnam, we can know that both are true of the system!

His conclusion is twofold. First, that a system may be described by not just one quantum state but two (or more). In some sense, each of these quantum states provides a true description of the system. Second, that these two quantum states need not be quantum-logically compatible.

Both of these claims are anathema to standard accounts of quantum theory. But quite apart from that, the results appear to run contrary to the whole quantum-logical approach. If $|\psi_0\rangle$ truly describes S, and $|\psi_1\rangle$ is logically incompatible with $|\psi_0\rangle$, then one would think that $|\psi_1\rangle$ cannot truly describe S. But in an unpredictable twist, Putnam tries to bring quantum logic to the rescue here. Since $|\psi_0\rangle$ and $|\psi_1\rangle$ are incompatible, their quantum logical conjunction is logically false. Nonetheless, each of the conjuncts is true! The solution, Putnam opines, is that *each of a pair of propositions may be true but their conjunction false.* That is what quantum logic, together with Putnam's decision to let more than one wavefunction describe a system, inexorably entail. But how to make sense of a false conjunction with two true conjuncts?

After all, it still seems perfectly acceptable to have a connective—call it #—defined so that A#B is true just in case each of A and B are true. From the truth of each of the pair one can validly infer the truth of A#B, and from the truth of A#B one can infer A and one can infer B. In all other cases, A#B is not true. What is gained by denying the very possibility of such a connective, and insisting on the quantum-logical one instead?

Here is his explication of the significance of quantum conjunction:

In effect, not allowing ourselves to conjoin all the statements we know to be true means that we have what amount to two different kinds of conjunction: One amounts to asserting statements in two different "frames" as I shall call them (different Boolean sub-logics); and the other, for which we reserve the *and*-sign, is conjunction of statements which lie in a common frame.[23]

Note that Putnam reserves the *and*-sign exactly for the "conjunction" that does not respect classical logic and designates no symbol at all for classical conjunction. That is, Putnam intends to *replace* the classical & with the quantum one. But why not at least leave the classical '&' alone and introduce a new symbol for the quantum operator? And what, after all, does the quantum conjunction *mean*?

With regard to the classical conjunction, Putnam thinks that certain aspects of quantum theory *refute* classical logic. But we know

───────────────

[23] Putnam, "Quantum Mechanics", 212.

that that cannot be correct: given the usual truth-functional semantics of the classical connectives, we know that classical inferences are non-ampliative. If you require that A and B both be true in order that A & B be true, and that the truth A and B separately follow from the truth of A & B, then you have defined the classical conjunction willy-nilly. It is quite irrelevant what "frames" A and B come from. You can't derive any non-conjunctive claim after conjunction has been introduced in this way that you couldn't already derive without it.

It is precisely here that Putnam introduces a new and unheralded consideration: predictive value. So far, as is appropriate, we have been concerned with the truth of propositions and nothing else. We have seen that we can stipulate that A & B is true exactly when A is true and B is true without harm. But predictive value is clearly a different property than truth. Putnam's claim is then twofold: 1) When using quantum conjunction of propositions we are concerned not with their truth but their predictive value and 2) there are some pairs of propositions such that they cannot both simultaneously have predictive value.

To sum up the situation, and contrary to Putnam's convention about the use of the & sign, we will use & to mean classical conjunction and ∧ to mean "quantum conjunction". Now suppose that A and B are propositions such that their quantum conjunction A ∧ B is the zero proposition, a zero-dimensional subspace. The conventional quantum logic, such as Finklestein's concludes that A and B cannot both be true, and *a fortiori* cannot both be known to be true. In other words, A & B can both be true and known. But according to Putnam, sometimes A and B can both be true and be known to be true. So A & B is true and knowable. Nonetheless, the unacceptability of A ∧ B indicates that A and B cannot both have predictive value. If A makes good predictions, then B cannot, and vice versa.

In essence, Putnam proposes using the vocabulary of classical logic for the connectives of a different logic, the "logic or predictive value". So what does it mean that one true proposition about a system has predictive value and another does not?

Perhaps a slight variation of Putnam's example will help. Once again, a shutter briefly opens, which may or may not release a photon. The shutter hole is covered by a polarizing filter that allows only horizontally polarized light through. Some distance away is a second polarizer oriented at 45° from the first, and then finally a screen.

The shutter is opened and a flash occurs on the screen. What can we infer about the photon from this result?

First, as Putnam notes, we can infer that a photon was indeed released while the shutter was open at $t_0$. But more than that, we can infer that the photon passed the first polarizer, and was therefore horizontally polarized at that time. Because the photon made it all the way to the screen, we can infer that it passed the second polarizer at $t_1$, and thereafter was polarized at 45°. These conclusions are unassailable from any interpretive point of view. Passing each polarizer counts as a "measurement" of the photo's polarization in the corresponding direction.

But Putnam's personal, and rather idiosyncratic insistence that the interactions we call "measurements" simply reveal some pre-interaction property of the measured system yields another, unexpected conclusion.

It means that at the time $t_2$, just before it reached the second polarizer, the photon was *both* polarized in the horizontal direction (as confirmed by passing the first polarizer) *and* polarized in the 45° direction (as confirmed by passing the second). But these two states are quantum-mechanically incompatible! Both are true, but the quantum-conjunction of them is necessarily false.

Putnam's idea that measurements simply reveal pre-existing values without disturbing the system, though, cannot survive more experimentation. Adding yet another polarizer oriented in the 45° direction would not change the output data at all: the same number of photons reach the screen. In other words, every photon that passes the second polarizer will, in this case, pass the third, which is oriented in the same direction. However, if the third polarizer is oriented in the horizontal direction, the number of photons will be reduced by half. Apparently, even though every photon that reaches the second polarizer is horizontally polarized, not every photon that passes the second polarizer is. The second polarizer can at least disturb the *horizontal* polarization, even if, as Putnam insists, it leaves the 45° polarization intact.

This effect of the second polarizer even appears somewhat startling if we change the third polarizer from horizontal to vertical. Now, with the second polarizer *removed*, no photons at all get through: every photon that passes the first polarizer is horizontally polarized and therefore is absorbed by the vertical polarizer. But re-inserting the 45° polarizer between these two—adding yet another filter between the two that are already blocking all the light—suddenly allows some light to get through! The beam is only 1/8 the brightness of the beam with no polarizers at all, but more light gets through three than just two in the same sequence. Obviously, the middle polarizer alters the photons that pass it.

The idea that "measurements" simply *reveal* pre-interaction values rather than *create* post-interaction values is, indeed, the hallmark of Putnam's take on quantum theory:

> Although the interpretation I have suggested is not realistic in the sense of assuming a copy theory of truth (metaphysical realism), or even in the sense of assuming that all observables have determinate values, it is *internally realistic* in the sense that *within the interpretation* no distinction appears between "measured values" and "unmeasured values".[24]

But how can this picture be even internally *coherent*? Suppose that a system *has* no determinate value for an observable, but is subject to a "measurement interaction" for that observable. Putnam's dictum seems to imply that *there can be no outcome at all*. For an outcome would assign a "measured value", which would clearly be distinct from the pre-interaction *lack* of any value.

The only way that "quantum logic" could come to the rescue here is to insist that although the pre-measurement system *had* no determinate value, and the post-measurement system *does have* a determinate value, you are not *allowed* to say both of these truths together, and draw the obvious conclusions from them, because they don't belong to "the same frame".  And this is, indeed, the sort of thing that some quantum logicians do say. But such a position is, to put it bluntly, madness.  If you can't always talk about the pre- and post-measurement situations in the same breath, then you can't always assert that there is no distinction between the values in the two cases, which is exactly what Putnam wants to say!

---

[24] Putnam, "Quantum Mechanics", 217-218.

And anyway, what would this prohibition on drawing conclusions from pairs of true sentences even have to do with conjunction? No need to conjoin A and B to infer from the premise set {"The pre-measurement system had no value for O.", "The post-measurement system had a value for O."} the conclusion "The pre-measurement state differs from the post-measurement state with respect to the value of O."

In short, Putnam's foray into his own version of quantum logic ended in self-contradiction and failure.

Putnam did come to recognize this. By the time of his his reply to Michael Redhead in *Reading Putnam*,[25] he had abandoned quantum logic, never to take it up again. He came to see that one can't just stipulate how so-called "measurement interactions" work and then trim logic itself to fit. The issue facing quantum theory is bringing "measurement interactions" and "measurements" under the same physical laws and principles of physical analysis as everything else.  What is left, then, is the legacy of a research programme that did not pan out. There are always things to learn from such projects, insights about the conditions for success and failure of a strategy. The main moral here is that the analogy to the fate of Euclidean geometry was flawed from the beginning. Abstract geometry has a clear physical counterpart that it can be used to describe: space or space-time. But there is no corresponding physical counterpart of the logical connectives.

---

[25] Bob Hale and Peter Clark (eds.), *Reading Putnam* (Oxford: Blackwell, 1994).

Kant went wrong when he took something extramental—space and time—and tried to make them merely the forms of our intuition. Since space and time are not just in our minds, our theories of space-time structure can get negative feedback from experience. But logical structure is not like space-time structure: there is no physical counterpart. As a result, physics cannot be simplified or improved by changing logic. Putnam wandered for three decades in the labyrinth of quantum logic, but he finally found his way out again.